\documentclass[draftclsnofoot, onecolumn, journal]{IEEEtran}

\usepackage{amsfonts, amssymb, amsthm}
\usepackage{graphicx}
\usepackage{mathtools}
\usepackage[normalem]{ulem}
\usepackage{bm}
\usepackage{array}
\usepackage{pgf}
\usepackage[capitalize]{cleveref}
\usepackage{microtype}
\usepackage{cite}

\newtheorem{theorem}{Theorem}[section]

\newtheorem{proposition}[theorem]{Proposition}

\theoremstyle{remark}
\newtheorem{example}[theorem]{Example}

\newcolumntype{L}{>{$}l<{$}}
\newcolumntype{C}{>{$}c<{$}}
\newcolumntype{R}{>{$}r<{$}}

\usepackage{enumitem}
\setlist[enumerate]{leftmargin=.5in}
\setlist[itemize]{leftmargin=.5in}

\DeclareMathOperator{\chr}{char}

\DeclareMathOperator{\spn}{span}
\DeclareMathOperator{\ev}{ev}
\DeclareMathOperator{\supp}{supp}

\DeclareMathOperator{\RS}{RS}
\DeclareMathOperator{\GRS}{GRS}
\DeclareMathOperator{\Rep}{Rep}

\renewcommand{\phi}{\varphi}
\renewcommand{\P}{\mathbb{P}}
\renewcommand{\hat}{\widehat}

\newcommand{\F}{\mathbb{F}}

\newcommand{\noise}{{\scriptscriptstyle\mathrm{noise}}}

\author{%
\IEEEauthorblockN{Okko~Makkonen, David~A.~Karpuk, Camilla~Hollanti \\
}%
\IEEEauthorblockA{
    Department of Mathematics and Systems Analysis\\
    Aalto University, Finland%
}%
\thanks{%
This work was done in part while O.~Makkonen and C.~Hollanti were visiting the Simons Institute for the Theory of Computing at the University of California, Berkeley. Preliminary results were presented at the 2024 IEEE International Symposium on Information Theory \cite{makkonen2024isit}.
This work has been supported by the Research Council of Finland under Grant No.\ 336005. The first author's work has been supported by the Vilho, Yrj\"o and Kalle V\"ais\"al\"a Foundation of the Finnish Academy of Science and Letters. Emails: okko.makkonen@aalto.fi, davekarpuk@gmail.com, camilla.hollanti@aalto.fi.
}}

\title{Secret Sharing for Secure and Private Information Retrieval: A Construction Using Algebraic Geometry Codes}

\begin{document}

\maketitle

\begin{abstract}
Private information retrieval (PIR) considers the problem of retrieving a data item from a database or distributed storage system without disclosing any information about which data item was retrieved. Secure PIR complements this problem by further requiring the contents of the data to be kept secure. Privacy and security can be achieved by adding suitable noise to the queries and data using methods from secret sharing. In this paper, a new framework for homomorphic secret sharing in secure and private information retrieval from colluding servers is proposed, generalizing the original cross-subspace alignment (CSA) codes proposed by Jia, Sun, and Jafar. We utilize this framework to give a secure PIR construction using algebraic geometry codes over hyperelliptic curves of arbitrary genus. It is shown that the proposed scheme offers interesting tradeoffs between the field size, file size, number of colluding servers, and the total number of servers. When the field size is fixed, this translates in some cases to higher retrieval rates than those of the original scheme. In addition, the new schemes exist also for some parameters where the original ones do not.
\end{abstract}

\section{Introduction}\label{sec:introduction}

Private Information Retrieval (PIR) \cite{chor1995private} studies the problem of retrieving a file from a database or distributed storage system without disclosing any information on the identity of the retrieved item. The basic variant of the problem considers public files, though security can be added to protect the contents of the files in addition to protecting user privacy. Scheme constructions for various scenarios with related capacity results can be found in the literature, e.g., \cite{Sun2016,sun2017repcap,Banawan2018,freij2017private,Tajeddine2018,freij2018tit,Oliveira2018isit,holzbaur2022tit}, including some quantum extensions \cite{ulukus_quantum_secure_pir, QTPIR}.

Cross-Subspace Alignment (CSA) codes have been recently proposed as a means to construct secure and private information retrieval schemes \cite{Jia_Sun_Jafar_XSTPIR,Jia_Jafar_MDSXSTPIR}. The necessary noise introduced to mask files and queries inevitably leads to the presence of \emph{interference} in the decoding process. CSA codes, as well as many other codes designed for various PIR problems, decompose the ambient space into a direct sum of an `information space' and a `noise space', and the goal of the scheme construction is mitigate the interference by minimizing the dimension of the latter under the constraint that the desired information is decodable.

In this work, we reinterpret the original CSA codes from \cite{Jia_Sun_Jafar_XSTPIR} as evaluation codes over the projective line, utilizing the framework of homomorphic secret sharing. In addition to a more conceptual construction, the resulting algebraic-geometric interpretation admits generalization to higher-genus curves. We focus on Algebraic Geometry (AG) codes on hyperelliptic curves of arbitrary genus, and showcase their potential in attaining PIR rates higher than the original CSA codes for a fixed field size by allowing a small increase in the number of servers and subpacketization. This improvement stems from the fact that increasing the genus yields curves with more rational points, which allows for longer code constructions. This allows for interesting tradeoffs between the field size, subpacketization, number of colluding servers, and the number of servers in total. Lowering the field size is of particular interest in applications where the communication bandwidth, computational capability or memory capacity of the storage nodes is limited, as is widely recognized in the literature evolving around codes for distributed data storage \cite{hou2016basic,raviv2017small}. Furthermore, some applications may operate on a very large number of storage nodes, and hence requiring the field size to grow with the number of nodes can become a limiting factor.

Let us also note that while this paper concentrates on the application of CSA codes to PIR, the AG framework is much more general and can be extended to cover several further instances of interference alignment in distributed systems. Namely, CSA codes have also been used as a means to construct Secure Distributed Matrix Multiplication (SDMM) schemes over classical and quantum channels \cite{Chen_Jia_Wang_Jafar_NGCSA,Jia_Jafar_SDMM,nsumboxarxiv,QCSA23} and for secure multi-party computation \cite{Chen_Jia_Wang_Jafar_NGCSA}. For some previous works on SDMM utilizing AG codes, we refer to \cite{machado2023hera,hollanti2023algebraic}. The framework introduced in the current paper should also serve as an invitation to revisit the star product PIR scheme \cite{freij2017private} and all of its subsequent derivatives, when the schemes are instantiated using algebraic geometry (AG) codes. With the exception of the preprint \cite{AGPIRarxiv}, wherein the author observes that the framework of \cite{freij2017private} is suitable for AG codes, the current authors know of no previous attempt at using AG codes for PIR, excluding the base case of Reed--Solomon codes. 

The rest of the paper is organized as follows. In \cref{sec:preliminaries}, the necessary mathematical background is outlined. \Cref{sec:secret_sharing} explains the framework of homomorphic secret sharing underlying any PIR scheme. Basics in PIR are given in \cref{sec:secure_and_private_information_retrieval}. \Cref{sec:projective_line_construction,sec:hyperelliptic_curve_construction} contain the detailed scheme constructions over the projective line and hyperelliptic curves, respectively. \Cref{sec:discussion} provides comparison of our construction to the CSA construction in \cite{Jia_Sun_Jafar_XSTPIR} and summarizes the work and provides directions for future research.

\section{Preliminaries}\label{sec:preliminaries}

In this section, we provide the background in coding theory and algebraic geometry relevant to our scheme constructions in \cref{sec:projective_line_construction,sec:hyperelliptic_curve_construction}.  The current work is meant to appeal to algebraic geometers who might be interested in new applications, as well as information or coding theorists unfamiliar to techniques from algebraic geometry.  As such, we provide the necessary background from both fields.

\subsection{Divisors and the Riemann--Roch Theorem}

We assume basic familiarity with the theory of algebraic curves, divisors, and Riemann--Roch spaces, but review crucial details in this subsection.  We let $\F_q$ denote the finite field with $q$ elements.  By a \emph{curve} over $\F_q$ we will always mean a smooth, projective, geometrically connected algebraic curve.

Let $\mathcal X$ be a curve over $\F_q$ with function field $\F_q(\mathcal X)$. Recall that a \emph{divisor} on $\mathcal X$ is a formal sum $D = \sum_{P\in\mathcal X}n_PP$ where all but finitely many $n_P = 0$. For a nonzero rational function $f \in \F_q(\mathcal{X})^*$ and a point $P \in \mathcal{X}$ we define $v_P(f) \in \mathbb{Z}$ to be the order of vanishing of $f$ at $P$. The divisor $(f) = \sum_{P\in\mathcal X}v_P(f)P$ of any nonzero $f \in \F_q(\mathcal{X})^*$ can be written as
\begin{equation*}
    (f) = (f)_0 - (f)_\infty,
    \quad\text{where}\quad
    (f)_0 = \!\! \sum_{P, v_P(f)>0} \!\! v_P(f)P 
    \quad\text{and}\quad
    (f)_\infty = \!\! \sum_{P, v_P(f)<0} \!\! -v_P(f)P
\end{equation*}
are the \emph{zero divisor} and \emph{pole divisor} of $f$, respectively. Moreover, we have $\deg((f)) = 0$.  Divisors of functions are also known as \emph{principal} divisors.

If $D$ is any divisor on $\mathcal X$, we can define the associated \emph{Riemann--Roch space}
\begin{equation*}
    \mathcal{L}(D) = \{ f \in \F_q(\mathcal{X}) \mid (f) + D \geq 0 \} \cup \{0\}.
\end{equation*}
This is a finite-dimensional vector space over $\F_q$, whose dimension is denoted by $\ell(D)$. If $\mathcal L(D)$ and $\mathcal L(D')$ are two such spaces, we can define their product to be
\[
\mathcal L(D)\cdot \mathcal L(D') = \spn_{\F_q}\{f\cdot g \mid f\in\mathcal L(D), g\in \mathcal L(D')\}.
\]
We define the \emph{maximum} of two divisors $D = \sum_{P\in \mathcal X}n_PP$ and $D' = \sum_{P\in\mathcal X}n_P'P$ to be the divisor $\max\{D, D'\} = \sum_{P \in \mathcal{X}} \max\{n_P, n'_P\} P$.

In the following theorem we collect all of the standard results about divisors that we require in the sequel.

\begin{theorem}\label{thm:riemann-roch_spaces}
Let $\mathcal{X}$ be a curve of genus $g$ over $\F_q$, and let $D, D'$ be divisors on $\mathcal{X}$.
\begin{enumerate}[label=\arabic*)]
    \item If $D$ is a divisor with $\deg(D) < 0$, then $\ell(D) = 0$.
    \item If $D \leq D'$, then $\mathcal{L}(D) \subseteq \mathcal{L}(D')$. It follows that $\mathcal L(D) + \mathcal L(D')\subseteq \mathcal L(\max\{D,D'\})$.
    \item If $D' = D + (h)$ then we have an isomorphism of vector spaces $\mathcal{L}(D') \to \mathcal{L}(D)$ given by $f \mapsto fh$. In particular, if $D = -(h)$ then $\mathcal L(D) = \spn\{h\}$.
    \item \cite[Theorem 8]{couvreur2017cryptanalysis} We have $\mathcal{L}(D) \cdot \mathcal{L}(D') \subseteq \mathcal{L}(D + D')$ with equality if $\deg(D) \geq 2g$ and $\deg(D') \geq 2g + 1$.
    \item (Riemann--Roch Theorem) If $\deg(D) > 2g - 2$ then $\ell(D) = \deg(D) - g + 1$.
\end{enumerate}
\end{theorem}
With the exception of 4) and 5), the above facts are all straightforward to verify from the definitions. The last point is of course only a consequence of the full Riemann--Roch Theorem, but it is all we require.

\subsection{Hyperelliptic Curves}

Here we review the required basics on hyperelliptic curves. We largely follow \cite[Chapter 13]{washington} and refer to this text for proofs of the results presented below.  By a \emph{hyperelliptic curve} over $\F_q$ we shall mean a curve $\mathcal X$ over $\F_q$ with affine equation
\[
y^2 + H(x)y = F(x)
\]
where $F$ is monic of $\deg(F) = 2g + 1$ and $\deg(H)\leq g$, for some $g\geq 1$.  It can be shown that $\mathcal X$ has a single point $P_\infty = [0:1:0]$ at infinity.  In case $\chr(\F_q)\neq 2$ one may assume $H = 0$.  Under these conditions, one can show that the genus of $\mathcal X$ is equal to $g$. More general models of hyperelliptic curves are possible \cite[Chapter 10]{galbraith}, but the above suffices for our purposes. When $g=1$, we call a hyperelliptic curve an \emph{elliptic curve}.

Now fix a hyperelliptic curve $\mathcal X$ of genus $g$ over $\F_q$ where $\chr(\F_q)\neq 2$, and consider the divisor $(k+g-1)P_\infty$ for some $k > g - 1$. By the Riemann--Roch theorem, we have $\ell((k+g-1)P_\infty) = k$. A basis for the Riemann--Roch space is given by
\begin{align*}
    \mathcal{L}((k + g - 1)P_\infty) &= \spn_{\F_q}\{ x^i y^j \mid 0 \leq i, 0 \leq j \leq 1, 2i + (2g + 1)j \leq k + g - 1 \} \\
    &= \spn_{\F_q}\{1, x, \ldots, x^{\lfloor (k + g - 1) / 2 \rfloor}, y, yx, \ldots, yx^{\lfloor (k - g - 2) / 2 \rfloor} \}.
\end{align*}
In particular, if $k \equiv g \pmod 2$, then
\begin{align*}
    \mathcal{L}((k + g - 1)P_\infty) = \spn_{\F_q}\{1, x, \ldots, x^{(k + g - 2) / 2}, y, yx, \ldots, yx^{(k - g - 2) / 2} \}.
\end{align*}

For a given genus $g$ and field size $q$, the \emph{Hasse--Weil bound} asserts that for any curve of genus $g$ over $\F_q$, we have
\[
    \lvert \mathcal X(\F_q) \rvert \leq q + 1 + 2g\sqrt{q}
\]
and a curve is \emph{maximal} if it achieves this bound. Determining when maximal curves exist for various values of $g$ and $q$ is an active area of research; see e.g.~\cite{garcia}.  On the other hand, given a hyperelliptic curve $\mathcal X$, we can define the \emph{hyperelliptic involution} $\iota \colon \mathcal X \to \mathcal X$ by the affine formula
\[
\iota(x,y) = (x, -y - H(x)).
\]
The map $\phi \colon \mathcal X \to \P^1$ defined on affine patches by $\phi(x,y) = x$ is a degree two rational map with fibers $\{P, \iota(P)\}$ for $P\in \mathcal X$. This gives the upper bound $\lvert \mathcal{X}(\F_q)\rvert \leq 2q + 1$. We will be particularly interested in hyperelliptic curves with many rational points, but this upper bound shows that we cannot hope to increase the number of points without bound by simply increasing the genus.

\subsection{Coding Theory Preliminaries}

An $[n,k,d]$ (linear) \emph{code} $\mathcal C$ is an $\F_q$-subspace of $\F_q^n$ with dimension $k$ and minimum (Hamming) distance $d$. We will omit the $d$ if unnecessary, and refer to $\mathcal C$ as an $[n,k]$ code. If $c\in \mathcal C$ we will refer to $c$ as a \emph{codeword}, and write its $i$th coordinate as $c(i)$. More generally, if $\mathcal T\subseteq [n]$ is any subset of coordinates, we denote by $c(\mathcal T)$ the projection of a codeword $c$ onto the coordinates in $\mathcal T$, and similarly by $\mathcal C(\mathcal T)$ the projection of the subspace $\mathcal C$ onto these coordinates. Any $\mathcal T\subseteq[n]$ of size $k$ such that $\dim(\mathcal C(\mathcal T)) = k$ is called an \emph{information set} of $\mathcal C$.

The Singleton bound asserts that $d\leq n - k + 1$, and if this bound is achieved with equality we say that $\mathcal C$ is \emph{Maximum Distance Separable} (MDS).

We denote by $\mathcal C^\perp$ the \emph{dual code} of $\mathcal C$, that is,
\begin{equation*}
    \mathcal C^\perp = \{v \in \F_q^n \mid \langle c, v \rangle = 0~\text{for all}~c \in \mathcal C\},
\end{equation*}
where $\langle \cdot, \cdot \rangle$ denotes the standard inner product on $\F_q^n$. The code $\mathcal C^\perp$ has length $n$ and dimension $n - k$. We denote its minimum distance by $d^\perp(\mathcal C)$ or simply $d^\perp$, and refer to this quantity as the \emph{dual minimum distance} of $\mathcal C$.

\subsection{Generalized Reed--Solomon Codes and Star Products}

Many PIR schemes are explicitly \cite{freij2017private} or, as we will demonstrate, implicitly \cite{Jia_Sun_Jafar_XSTPIR} based on \emph{Generalized Reed--Solomon} (GRS) codes and their \emph{star products}. Let us briefly review the construction of such codes and their behavior with respect to the star product of two linear codes.

For any $k > 0$ and any polynomial $0 \neq f \in \F_q[x]$ we define the $\F_q$-subspaces
\begin{equation*}
    \F_q[x]^{<k} \coloneqq \{p\in \F_q[x] \mid \deg(p) < k\} \quad \text{and} \quad f \cdot \F_q[x]^{<k} \coloneqq \{f \cdot p \mid p \in \F_q[x]^{<k}\}
\end{equation*}
of the polynomial ring $\F_q[x]$, each of which clearly has dimension $k$.

To construct a GRS code of length $n$ and dimension $k\leq n$, we choose a vector $\alpha = (\alpha_1,\ldots,\alpha_n)\in\F_q^n$ of distinct evaluation points, and a polynomial $f\in \F_q[x]$ such that $f(\alpha_i)\neq 0$ for all $i$. Let $\nu = (f(\alpha_1),\ldots,f(\alpha_n))$. We consider the \emph{evaluation map}
\begin{equation*}
    \ev_\alpha \colon \F_q[x]\to \F_q^n,\quad \ev_\alpha(h) = (h(\alpha_1), \ldots, h(\alpha_n))
\end{equation*}
and define $\GRS_k(\alpha,\nu)$ as the image of $f\cdot \F_q[x]^{<k}$ under this map. It is straightforward to verify that $\GRS_k(\alpha,\nu)$ is an MDS code of length $n$ and dimension $k$. If $f$ is a nonzero constant polynomial, the resulting code is simply called a \emph{Reed--Solomon code} and denoted by $\RS_k(\alpha)$.
 
The star product of two linear codes $\mathcal C$ and $\mathcal D$ of length $n$ over $\F_q$ is defined to be
\begin{equation*}
    \mathcal C \star \mathcal D \coloneqq \spn_{\F_q}\{c\star d \mid c\in \mathcal C, d\in \mathcal D\}
\end{equation*}
where $c\star d = (c(1)d(1),\ldots,c(n)d(n))$. Note that when we endow $\F_q^n$ with the star product, it becomes an $\F_q$-algebra in a natural way. Now if $V$ and $W$ are any finite-dimensional $\F_q$\nobreakdash-subspaces of $\F_q[x]$, we can define $V\cdot W = \spn_{\F_q}\{f\cdot g \mid f\in V, g\in W\}$. Letting $\mathcal C = \ev_\alpha(V)$ and $\mathcal D = \ev_\alpha(W)$, we arrive at the identity
\[
\ev_\alpha(V\cdot W) = \mathcal C\star \mathcal D
\]
which expresses the compatibility of the product of polynomials and the star product of their evaluation vectors.

When we choose $V = f\cdot \F_q[x]^{<k}$ and $W = g\cdot \F_q[x]^{<\ell}$, that is, let the above $\mathcal C$ and $\mathcal D$ be GRS codes, we have that $V\cdot W = fg\cdot \F_q[x]^{<k+\ell-1}$. Applying the map $\ev_\alpha$ to both sides of this equation yields the star product identity
\[
\GRS_k(\alpha, \nu)\star \GRS_\ell(\alpha, \mu) = \GRS_{\min\{k+\ell-1,n\}}(\alpha, \nu\star\mu)
\]
where $\nu = (f(\alpha_1),\ldots,f(\alpha_n))$ and $\mu = (g(\alpha_1),\ldots,g(\alpha_n))$.

\subsection{Algebraic Geometry Codes}

One generalizes the construction of GRS codes to \emph{Algebraic Geometry} codes, or simply AG codes, as follows. We provide \cite{hoholdt} as a catch-all reference for this topic.

We fix a curve $\mathcal X$ over $\F_q$ with genus $g$ and rational function field $\F_q(\mathcal X)$. Let $\mathcal S$ be a non-empty, finite subset of $\mathcal X$ and set $U = \mathcal X \setminus \mathcal S$, which is Zariski open. We let $\mathcal P = \{P_1,\ldots,P_n\}\subseteq U$ be a set of $\F_q$-rational \emph{evaluation points} on $\mathcal X$. We consider the ring $\mathcal O_{\mathcal X}(U)\subseteq \F_q(\mathcal X)$ of regular functions on $U$, and the evaluation map
\[
\ev_{\mathcal P} \colon \mathcal O_{\mathcal X}(U) \to \F_q^n, \quad \ev_{\mathcal P}(h) = (h(P_1),\ldots,h(P_n)).
\]
If $D$ is a divisor on $\mathcal X$ such that $\supp(D) \subseteq \mathcal S$, then we define the AG code $\mathcal C(\mathcal P, D)$ to be the image of $\mathcal L(D)$ under this map.

Assuming that $n > \deg(D)$, then $\mathcal C(\mathcal P, D)$ is an $[n,k,d]$ code with $k = \ell(D)$ and $d\geq n - \deg(D)$. By \cref{thm:riemann-roch_spaces}, if $\deg(D) > 2g - 2$, then $\ell(D) = \deg(D) - g + 1$, so combining with the Singleton bound we get
\begin{equation}\label{eq:AG_singleton_bound}
    n - k + 1 - g \leq d \leq n - k + 1.
\end{equation}
Using the well-known fact that the dual of an AG code is again an AG code, one can derive the following bound on the dual minimum distance $d^\perp$ of such a code:
\begin{equation}\label{eq:AG_dual_distance}
    k + 1 - g \leq d^\perp \leq k + 1
\end{equation}
From \eqref{eq:AG_singleton_bound} one sees that AG codes on curves with genus $g = 0$ are MDS codes. More precisely, we have the following example.

\begin{example}\label{ex:grs_example}
Let $\mathcal X = \mathbb P^1$ and let $P_\infty = [0:1]$ be the point at infinity. Consider a divisor $D$ with $\deg(D) \geq 0$ and let $k = \deg(D) + 1$. Since every degree zero divisor on $\P^1$ is principal, we may write $D = (k - 1)P_\infty - (h)$ for some function $h \in \F_q(x)^*$. Letting $P_i = {[\alpha_i : 1]} \in \P^1$, setting $\nu_i = h(\alpha_i)$ and assuming $\nu_i \neq 0$, one has $\mathcal{C}(\mathcal{P}, D) = \GRS_k(\alpha, \nu)$. This means that the AG codes over the projective line are exactly the GRS codes. In particular, if $D = 0$, then $\mathcal{C}(\mathcal{P}, D) = \Rep(n)$, the length $n$ repetition code.
\end{example}

If $V$ and $W$ are finite-dimensional $\F_q$-subspaces of $\mathcal O_{\mathcal X}(U)$ then we can define $V\cdot W = \spn_{\F_q}\{f\cdot g \mid f\in V, g\in W\}\subseteq \mathcal O_{\mathcal X}(U)$. Setting $V = \mathcal L(D)$ and $W = \mathcal L(D')$ for some divisors $D$ and $D'$ whose supports are both contained in $\mathcal S$, we have
\[
\mathcal L(D)\cdot \mathcal L(D') \subseteq \mathcal L(D+D') \subseteq \mathcal O_{\mathcal X}(U)
\]
with equality in the first containment if $\deg(D) \geq 2g$ and $\deg(D')\geq 2g+1$ by \cref{thm:riemann-roch_spaces}. Applying the map $\ev_{\mathcal P}$ to both sides of the first containment we arrive at the identity
\[
\mathcal C(\mathcal P, D)\star \mathcal C(\mathcal P, D') \subseteq \mathcal C(\mathcal P, D + D')
\]
where we again have equality under the previously mentioned assumptions.

\section{Secret Sharing}\label{sec:secret_sharing}

An encryption scheme turns a message $m$ into a ciphertext $c$ using a random transformation.  We say an encryption scheme has \emph{perfect security} if the distribution of the ciphertext does not depend on the message $m$. One way to achieve perfect security is by using a \emph{one-time pad}.  In particular, let $G$ be a finite additive group and let $m \in G$ be a message. Choose $r \in G$ uniformly at random and set $c = m + r$. It is easy to verify that $c$ is uniformly distributed and independent of $m$ for all $m$, so the one-time pad has perfect security. 

\emph{Secret sharing} is a way of distributing a secret value to $N$ parties such that only some admissible sets of parties are able to recover the secret, while some forbidden sets of parties will not be able to deduce anything about the secret. The security in secret sharing is often based on the one-time pad, which provides perfect security against any forbidden subset of the parties.  One can view such a secret sharing scheme as a collection of `local' one-time pads, where the noise one adds is not uniform on the ambient space, but appears uniform when we restrict our attention to any forbidden subset.

\subsection{Linear Secret Sharing}

One way to construct secret sharing schemes from linear codes is given as follows.  We begin by choosing two linear codes $\mathcal{C}, \mathcal{C}^\noise \subseteq \F_q^N$. The secret value is encoded as a codeword $c \in \mathcal{C}$ and the secret share $\hat{c}$ is chosen uniformly at random from $c + \mathcal{C}^\noise$. Each party $n \in [N]$ is then given the coordinate $\hat{c}(n)$. The secret value should be uniquely decodable from all the shares $\hat{c}$, so we require that $\mathcal{C} \cap \mathcal{C}^\noise = 0$, which makes the sum $\hat{\mathcal{C}} = \mathcal{C} + \mathcal{C}^\noise$ direct. The pair of codes $(\mathcal{C}, \mathcal{C}^\noise)$ that satisfies $\mathcal{C} \cap \mathcal{C}^\noise = 0$ is said to be a \emph{linear secret sharing scheme}, or LSSS, over the field $\F_q$. We say that a scheme $(\mathcal{C}, \mathcal{C}^\noise)$ is \emph{$T$-secure}, if the distribution of any subset of $T$ shares does not depend on the secret, i.e., provides perfect security for the secret. For a general reference on LSSS's see \cite[Section 4.2]{chen2007secure}.

A set of $T$ compromised parties $\mathcal{T} \subseteq [N]$ will observe the projection $\hat{c}(\mathcal{T}) \in c(\mathcal{T}) + \mathcal{C}^\noise(\mathcal{T})$. If $d^\perp(\mathcal{C}^\noise) > T$, then any $T$ columns of the generator matrix of $\mathcal{C}^\noise$ will be linearly independent, so $\mathcal{C}^\noise(\mathcal{T}) = \F_q^T$. Therefore, $\hat{c}(\mathcal{T}) \in c(\mathcal{T}) + \F_q^T$ is uniformly distributed, which means that the compromised parties will observe $c(\mathcal{T})$ one-time padded with uniform noise. We formulate this as the following standard result about LSSS's.

\begin{proposition}\label{prop:secret_sharing_security}
An LSSS $(\mathcal{C}, \mathcal{C}^\noise)$ is $T$-secure for $T = d^\perp(\mathcal{C}^\noise) - 1$.
\end{proposition}

\begin{example}[Shamir secret sharing]\label{ex:shamir_secret_sharing}
Secret sharing was first introduced by Shamir in 1979 \cite{shamir79}. This construction fits in the linear secret sharing framework described above as the codes $\mathcal{C}$ and $\mathcal{C}^\noise$ are chosen to be GRS codes. In particular, we choose the polynomial $\hat{f} \in f + x \cdot \F_q[x]^{< T}$ uniformly at random, where $f \in \F_q$ is the secret. The shares are chosen as $\hat{c}(n) = \hat{f}(\alpha_n)$, where $\alpha_1, \dots, \alpha_N$ are distinct elements in $\F_q$. This corresponds to the codes
\begin{equation*}
    \mathcal{C} = \ev_\alpha(\F_q) = \RS_1(\alpha) \quad \text{and} \quad \mathcal{C}^\noise = \ev_\alpha(x \cdot \F_q[x]^{< T}) = \GRS_T(\alpha, \alpha),
\end{equation*}
where $\F_q \subseteq \F_q[x]$ is identified with the constant polynomials. As long as $\alpha_n \neq 0$ for all $n \in [N]$, we have that $d^\perp(\mathcal{C}^\noise) = T + 1 > T$. Instead of codes and codewords, we will often work with function spaces and functions with the assumption that we have an evaluation map that can turn these into codewords in some suitable code.
\end{example}

\begin{example}[Chen--Cramer secret sharing]\label{ex:chen_cramer_secret_sharing}
Let $\mathcal X$ be a curve over $\F_q$ of genus $g$. The Chen--Cramer secret sharing scheme of \cite{chen_ag_secret_share} generalizes the Shamir scheme as follows. Let $P_\infty$ be a fixed rational point of $\mathcal X$ and consider the divisor $(T + 2g - 1)P_\infty$ for some security parameter $T>0$, let $h\in \F_{q}(\mathcal X)^*$ be nonconstant and have zero divisor disjoint from $P_\infty$, and choose a set $\mathcal P = \{P_1,\ldots,P_N\}\subseteq\mathcal X(\F_q)$ of evaluation points avoiding $P_\infty$ and $(h)$. Suppose we have a secret $f\in \F_{q}$. Choose a rational function $\hat{f}\in f + h\cdot \mathcal L((T + 2g - 1)P_\infty)$ uniformly at random. The shares are now chosen to be $\hat{c}(n) = \hat{f}(P_n)$. This corresponds to the codes
\[
\mathcal C = \ev_{\mathcal P}(\F_{q})
\quad\text{and}\quad
\mathcal C^\noise = \mathcal C(\mathcal P, (T + 2g - 1)P_\infty-(h)).
\]
Clearly any function in $h\cdot \mathcal L((T + 2g - 1)P_\infty)$ has zeros at the zeros of $h$, hence the two codes intersect trivially. Assuming that $N > T + 2g - 1$ so that $\dim(\mathcal C^\noise) = T+g$, the LSSS $(\mathcal C, \mathcal C^\noise)$ is $T$-secure.   

For example, consider the (maximal) hyperelliptic curve $\mathcal X$ of genus $g=2$ defined over $\F_{13}$ by the affine equation
\[
y^2 = x^5+x^4+4x^2+2x+1
\]
which has a single point $P_\infty$ at infinity and $|\mathcal X(\F_{13})| - 1 = 25$ non-infinite rational points. This curve was obtained from the tables at \cite{many_points}. Set $h = y$, which has a single rational zero at $Q = (5,0)$. Let $\mathcal P = \mathcal X(\F_{13})\setminus\{P_\infty, Q\}$. Setting the security parameter to be $T = 4$, we have
\[
\mathcal L((T + 2g - 1)P_\infty) = \mathcal L(7P_\infty) = \spn_{\F_{13}}\{1, x, x^2, x^3, y, xy\}.
\]
The code $\mathcal C^\noise = \mathcal C(\mathcal P, 7P_\infty - (y))$ has parameters $[24, 6, 17]$ and dual minimum distance $d^\perp(\mathcal C^\noise) = T + 1 = 5$, and therefore the LSSS $(\mathcal C, \mathcal C^\noise)$ is $4$-secure.  Note that defining the Shamir scheme for $N = 24$ parties and security parameter $T\geq 4$ would require $q\geq 25$. 
\end{example}

\subsection{Homomorphic Secret Sharing}

Instead of just one secret value, assume that the parties hold the secret shares of $L$ different secrets and we wish to decode a linear combination of the secrets. We could do this by downloading all of the shares individually and computing the linear combination of the secrets. However, doing this would come at a large bandwidth cost as we download $L$ shares. Instead, we can download a linear combination of the shares and decode the linear combination of the secrets directly. If $\hat{c}_\ell \in c_\ell + \mathcal{C}^\noise$, then we receive
\begin{equation*}
    \sum_{\ell \in [L]} \lambda_\ell \hat{c}_\ell \in \sum_{\ell \in [L]} \lambda_\ell c_\ell + \mathcal{C}^\noise,
\end{equation*}
for some coefficients $\lambda_1, \dots, \lambda_L \in \F_q$. This means that the result is a secret share of the linear combination in the secret sharing scheme $(\mathcal{C}, \mathcal{C}^\noise)$.

Instead of decoding a linear combination of the secrets, we may want to decode all of the secrets individually. Again, we could do this by downloading all of the shares individually and incurring a larger bandwidth cost. Instead, we download the sum of the shares from each party. If $\hat{c}_\ell \in c_\ell + \mathcal{C}_\ell^\noise$, then we observe
\begin{equation*}
    \sum_{\ell \in [L]} \hat{c}_\ell \in \sum_{\ell \in [L]} c_\ell + \sum_{\ell \in [L]} \mathcal{C}_\ell^\noise,
\end{equation*}
where $c_\ell \in \mathcal{C}_\ell$. In this case the individual secret sharing schemes $(\mathcal{C}_\ell, \mathcal{C}_\ell^\noise)$ may consist of different linear codes. However, the sum is a secret share in the LSSS defined by the codes $\mathcal{C} = \mathcal{C}_1 + \dots + \mathcal{C}_L$ and $\mathcal{C}^\noise = \mathcal{C}_1^\noise + \dots + \mathcal{C}_L^\noise$. Then, $\sum_\ell c_\ell$ is decodable from the sum if and only if $\mathcal{C} \cap \mathcal{C}^\noise = 0$. Furthermore, the individual secret values $c_1, \dots, c_L$ will be decodable from $\sum_\ell c_\ell$ if and only if the subspaces $\mathcal{C}_1, \dots, \mathcal{C}_L$ are linearly independent, i.e., $\dim(\sum_\ell \mathcal{C}_\ell) = \sum_\ell \dim(\mathcal{C}_\ell)$.

\begin{example}\label{ex:homomorphic_secret_sharing}

We continue with \cref{ex:shamir_secret_sharing}.  Let $f_1, f_2 \in \F_q$ be two secrets that are secret shared using the polynomials $\hat{f}_1 \in f_1 + x \cdot \F_q[x]^{< T}$ and $\hat{f}_2 \in f_2 + x \cdot \F_q[x]^{< T}$. Then their sum will be
\begin{equation*}
    \hat{f}_1 + \hat{f}_2 \in (f_1 + f_2) + x \cdot \F_q[x]^{< T}.
\end{equation*}
As $f_1 + f_2$ is the constant term, we can not decode both $f_1$ and $f_2$, but only their sum. On the other hand, if we modify the secret shares to be $\hat{f}_1 \in f_1 + x^2 \cdot \F_q[x]^{< T}$ and $\hat{f}_2 \in f_2 \cdot x + x^2 \cdot \F_q[x]^{< T}$, then the sum is
\begin{equation*}
    \hat{f}_1 + \hat{f}_2 \in f_1 + f_2x + x^2 \cdot \F_q[x]^{< T}.
\end{equation*}
Now, after passing to appropriate evaluation vectors, $f_1$ and $f_2$ can be decoded, since $1$ and $x$ are linearly independent.
\end{example}

Consider the case that the parties hold shares for two secrets and we want to obtain the product of these secret values. Decoding each secret individually is not practical, so we ask the parties to compute the product of their shares and return it. If the shares are $\hat{c} \in c + \mathcal{C}^\noise$ and $\hat{d} \in d + \mathcal{D}^\noise$ coming from LSSS's $(\mathcal{C}, \mathcal{C}^\noise)$ and $(\mathcal{D}, \mathcal{D}^\noise)$, then the product will be
\begin{equation*}
    \hat{c} \star \hat{d} \in c \star d + \mathcal{C} \star \mathcal{D}^\noise + \mathcal{C}^\noise \star \mathcal{D} + \mathcal{C}^\noise \star \mathcal{D}^\noise.
\end{equation*}
This is a secret share of $c \star d$ in the LSSS defined by
\begin{equation*}
    \mathcal{E} = \mathcal{C} \star \mathcal{D} \quad \text{and} \quad \mathcal{E}^\noise = \mathcal{C} \star \mathcal{D}^\noise + \mathcal{C}^\noise \star \mathcal{D} + \mathcal{C}^\noise \star \mathcal{D}^\noise.
\end{equation*}
Again, to be able to decode, we should have that these codes intersect trivially. To achieve this, the code $\mathcal{E}^\noise$ should not have a large dimension. However, the dimension of the star product of generic codes $\mathcal{C}^\noise, \mathcal{D}^\noise \subseteq \F_q^N$ will have dimension $\min\{N, \dim(\mathcal{C}^\noise)\dim(\mathcal{D}^\noise)\}$, which means that we need to choose these codes in a clever manner.

\section{Secure and Private Information Retrieval}\label{sec:secure_and_private_information_retrieval}

Consider a database consisting of $M$ files $s_1,\ldots,s_M\in \F_q^L$, stored in a distributed fashion across $N$ servers. We refer to a single coordinate of such a vector as a \emph{fragment} of a file. These fragments could themselves be vectors over $\F_q$, but for our purposes there is no loss of generality in simply assuming the fragments are scalars $s_{m, \ell}\in \F_q$, for $\ell \in [L]$ and $m\in [M]$. In private information retrieval (PIR), we want to retrieve one of these files $s_\mu = (s_{\mu, 1}, \ldots, s_{\mu, L})$ without revealing the desired index $\mu\in [M]$ to the servers. We will retrieve the file by sending \emph{queries} to the servers who combine them with their \emph{stored data} and send back their \emph{responses}. The precise nature of how the data is stored, how we query each server, and the method we use to recover the desired file from the responses, will soon be made precise.

To have \emph{data security}, we require that the data stored at any $X$ servers reveals nothing about the file contents. Similarly, to have \emph{query privacy}, we require that the queries sent to any $T$ servers reveal nothing about the desired file index. For a precise information-theoretic formulation of secure and private information retrieval, see \cite{Jia_Sun_Jafar_XSTPIR}.

For us, a \emph{PIR scheme} is a set of pairs of linear secret sharing schemes $(\mathcal{C}_\ell, \mathcal{C}_\ell^\noise)$ and $(\mathcal{D}_\ell, \mathcal{D}_\ell^\noise)$ over $\F_q$ for $\ell \in [L]$, each with $N$ parties. We will use the LSSSs $(\mathcal{C}_\ell, \mathcal{C}_\ell^\noise)$ to secret share the file fragments with the $N$ servers and the LSSSs $(\mathcal{D}_\ell, \mathcal{D}_\ell^\noise)$ to secret share the queries. We say that a PIR scheme is \emph{$X$-secure} if the LSSSs $(\mathcal{C}_\ell, \mathcal{C}_\ell^\noise)$ are $X$-secure. Similarly, we say that a PIR scheme is \emph{$T$-private} if the LSSSs $(\mathcal{D}_\ell, \mathcal{D}_\ell^\noise)$ are $T$-secure. It is easy to verify that the conditions in \cite{Jia_Sun_Jafar_XSTPIR} are satisfied by our definitions. The (download) \emph{rate} of a PIR scheme is defined as the ratio of the amount of desired information (the file size) and the amount of information downloaded while retrieving the file, which in our case is $\mathcal{R} = \tfrac{L}{N}$.  

For simplicity we assume that the codes $\mathcal{C}_\ell$ and $\mathcal{D}_\ell$ are one-dimensional and spanned by some vectors $c_\ell$ and $d_\ell$, respectively. Realizing the file fragments $s_{m, \ell}$ and the queries $q_{m, \ell}$ as elements of $\F_q$, we then secret share these objects among the $N$ servers according to
\begin{equation*}
    \hat{s}_{m, \ell} \in s_{m, \ell} c_\ell + \mathcal{C}_\ell^\noise, \quad \hat{q}_{m, \ell} \in q_{m, \ell} d_\ell + \mathcal{D}_\ell^\noise.
\end{equation*}
The server indexed by $n \in [N]$ stores $\hat{s}_{m, \ell}(n)$ for all $\ell \in [L]$ and $m \in [M]$. We query this server with $\hat{q}_{m, \ell}(n)$ and they respond with the inner product
\begin{equation*}
    r(n) = \sum_{\ell \in [L]} \sum_{m \in [M]} \hat{s}_{m, \ell}(n) \hat{q}_{m, \ell}(n).
\end{equation*}
Since we are free to pick the queries $q_{m, \ell}$ as we like, we set $q_{m, \ell} = \delta_{m, \mu}$, the Kronecker delta picking out the desired term in the sum.  Thus we can write a desired file fragment $s_{\mu, \ell}$ as the linear combination
\[
s_{\mu, \ell} = \sum_{m\in [M]}s_{m, \ell}q_{m, \ell}
\]
which demonstrates that this indeed is a particular instance of homomorphic secret sharing.  The total vector $r = (r(1),\ldots,r(N))$ we receive is given by
\begin{equation*}
    r = \sum_{\ell \in [L]} \sum_{m \in [M]} \hat{s}_{m, \ell} \star \hat{q}_{m, \ell} \in \sum_{\ell \in [L]} \bigg( \sum_{m \in [M]} s_{m, \ell}q_{m, \ell} \bigg) e_\ell + \mathcal{E}^\noise = \sum_{\ell\in[L]}s_{\mu, \ell}e_\ell + \mathcal E^\noise
\end{equation*}
where $e_\ell = c_\ell \star d_\ell$ and the \emph{noise space} $\mathcal E^\noise$ is defined to be
\begin{equation*}
    \mathcal{E}^\noise = \sum_{\ell \in [L]} \big( \mathcal{C}_\ell \star \mathcal{D}_\ell^\noise + \mathcal{C}_\ell^\noise \star \mathcal{D}_\ell + \mathcal{C}_\ell^\noise \star \mathcal{D}_\ell^\noise \big).
\end{equation*}
From this expression we must be able to decode $s_{\mu, \ell}$ for all $\ell \in [L]$ to recover the desired file.

Now define the subspaces $\mathcal E_\ell = \mathcal C_\ell \star \mathcal D_\ell = \spn\{e_\ell\}$, whose sum $\mathcal E = \sum_\ell \mathcal E_\ell$ we refer to as the \emph{information space}.  To be able to decode the $s_{\mu, \ell}$ we require that $\mathcal{E} \cap \mathcal{E}^\noise = 0$ and $e_\ell$ are linearly independent, i.e., $\dim(\mathcal{E}) = L$.   If this condition holds, then we see that $r$ is a secret share in the LSSS defined by $(\mathcal{E}, \mathcal{E}^\noise)$. We summarize the connection between homomorphic secret sharing schemes and $X$-secure $T$-private information retrieval in the following theorem.

\begin{theorem}\label{thm:PIR_from_secret_sharing}
Let $(\mathcal{C}_\ell, \mathcal{C}_\ell^\noise)$ and $(\mathcal{D}_\ell, \mathcal{D}_\ell^\noise)$ for $\ell \in [L]$ be linear secret sharing schemes over $\F_q$ with $N$ parties such that $\dim(\mathcal{C}_\ell) = \dim(\mathcal{D}_\ell) = 1$, $\mathcal{E} \cap \mathcal{E}^\noise = 0$, and $\dim(\mathcal{E}) = L$, where $\mathcal{E}, \mathcal{E}^\noise$ are defined as above. Then there exists a PIR scheme with rate $\mathcal R = \tfrac{L}{N}$ which is $X$-secure for $X = \min_\ell d^\perp(\mathcal{C}_\ell^\noise) - 1$ and $T$-private for $T = \min_\ell d^\perp(\mathcal{D}_\ell^\noise) - 1$.
\end{theorem}

The authors of \cite{Jia_Sun_Jafar_XSTPIR} construct an $X$-secure $T$-private information retrieval scheme with rate $\mathcal{R} = 1 - \tfrac{X + T}{N}$ and $N = L + X + T$ servers over a field of size $q \geq N + L = 2L + X + T$.  They also show that this rate is information-theoretically optimal as the number of files $M$ approaches infinity.

We remark that the above-described scheme carries through if the queries $q_{m, \ell}$ are any constants whatsoever.  In other words, the above construction generalizes in a straightforward way to compute any $L$ linear functions of a database while achieving the same rate.  

Let us conclude this subsection by making a connection with PIR schemes for \emph{coded} distributed storage systems \cite{Banawan2018}, wherein data is divided into blocks and then distributed among $N$ servers using a \emph{storage code}.  One could rephrase the coding done in the above construction to guarantee $X$-security as employing such a storage code, but then applying the methods for PIR for coded systems as in \cite{Banawan2018, freij2017private} results in one effectively decoding vectors of length $X+1$ to obtain a single scalar.  The net effect is the presence of an undesirable multiplicative factor of $\tfrac{1}{X+1}$ in the rate.  This strategy was employed in \cite{raviv_karpuk}, but clearly the scheme of \cite{Jia_Sun_Jafar_XSTPIR}, which does not suffer this defect, achieves superior performance.

\section{Construction Using the Projective Line}\label{sec:projective_line_construction}

In this section we reconstruct the CSA codes of \cite{Jia_Sun_Jafar_XSTPIR} as Algebraic Geometry codes on the projective line $\mathbb{P}^1$, i.e., GRS codes.  As a first step we demonstrate the construction solely in terms of polynomials (i.e.~regular functions on an affine patch of $\mathbb{P}^1$), and then interpret the construction geometrically. This geometric interpretation allows us to generalize these codes to higher genus curves.  In general, scheme constructions are often streamlined by working directly with polynomials and rational functions instead of the evaluation vectors they define.

The general framework of the previous two sections is well-suited for evaluation codes. In particular, we realize $\mathcal C_\ell$, $\mathcal C_\ell^\noise$, $\mathcal D_\ell$, $\mathcal D_\ell^\noise$ as GRS codes. The dictionary between GRS codes and $\F_q$\nobreakdash-subspaces of the form $h \cdot \F_q[x]^{<K}$, and therefore between codewords and polynomials, allows us to describe pairs of linear secret sharing schemes $(\mathcal C_\ell, \mathcal C_\ell^\noise)$ and $(\mathcal D_\ell, \mathcal D_\ell^\noise)$ for $\ell\in [L]$ which satisfy the hypotheses of \cref{thm:PIR_from_secret_sharing} solely in terms of polynomials.  

\subsection{Construction by Polynomials}

Let $h$ be a polynomial of degree $L$. Then
\begin{equation*}
    \F_q[x]^{< L} \oplus h \cdot \F_q[x] = \F_q[x]
\end{equation*}
due to the polynomial division algorithm in $\F_q[x]$. We want to utilize this direct sum decomposition in our construction. More specifically, for some $N> L$, we truncate the above direct sum decomposition to get
\begin{equation}
    \F_q[x]^{<L} \oplus h\cdot \F_q[x]^{< N-L} = \F_q[x]^{<N}  \label{eq:truncated_polynomial_division_decomposition}
\end{equation}
and the two summands on the left-hand side will play the role of $\mathcal E$ and $\mathcal E^\noise$, respectively. Our goal now is to use GRS codes to construct appropriate secret sharing schemes $(\mathcal C_\ell, \mathcal C_\ell^\noise)$ and $(\mathcal D_\ell, \mathcal D_\ell^\noise)$ so that the resulting pair $(\mathcal E, \mathcal E^\noise)$ as described previously section is as desired.

Let $f_\ell, g_\ell \in \F_q$ be scalars whose product we want to compute. Similar to \cref{ex:homomorphic_secret_sharing} we choose the secret shares to be
\begin{equation}\label{eq:projective_line_secret_shares}
    \hat{f}_\ell \in f_\ell + f_\ell^\noise \cdot \F_q[x]^{< X}, \quad \hat{g}_\ell \in g_\ell h_\ell + g_\ell^\noise \cdot \F_q[x]^{< T},
\end{equation}
where $h_\ell$, $f_\ell^\noise$ and $g_\ell^\noise$ are suitably chosen polynomials. The sum of the products of the secret shares is then
\begin{align*}
    \sum_{\ell \in [L]} \hat{f}_\ell \hat{g}_\ell &\in \sum_{\ell \in [L]} f_\ell g_\ell h_\ell + \sum_{\ell \in [L]}(g_\ell^\noise \cdot \F_q[x]^{< T} + f_\ell^\noise h_\ell \cdot \F_q[x]^{< X} + f_\ell^\noise g_\ell^\noise \cdot \F_q[x]^{< X + T - 1}).
\end{align*}
To take advantage of the direct sum decomposition of \eqref{eq:truncated_polynomial_division_decomposition}, we want the terms in the first sum to have degree $< L$ and be linearly independent, hence the presence of the polynomial $h_\ell$. Furthermore, we want the terms in the second sum to be multiples of $h$. To achieve this, we assert that (1) $h_\ell$ are a basis of $\F_q[x]^{< L}$, (2) $g_\ell^\noise = h$, and (3) $f_\ell^\noise = h/h_\ell$ (assuming that $h_\ell$ divides $h$). With these choices one can verify that
\begin{equation*}
    \sum_{\ell \in [L]} \hat{f}_\ell \hat{g}_\ell \in \sum_{\ell \in [L]} f_\ell g_\ell h_\ell + h \cdot \F_q[x]^{< X + T - 1 + \Delta},
\end{equation*}
where $\Delta = \max_\ell \deg(h/h_\ell)$. In particular, the above polynomial has degree $< L + X + T - 1 + \Delta$.

These choices for the polynomials mean that the corresponding secret sharing codes are
\begin{alignat*}{3}
    \mathcal{C}_\ell &= \ev_\alpha(\F_q), \quad &&\mathcal{C}_\ell^\noise &&= \ev_\alpha(h/h_\ell \cdot \F_q[x]^{<X}), \\
    \mathcal{D}_\ell &= \ev_\alpha(h_\ell \cdot \F_q), \quad &&\mathcal{D}_\ell^\noise &&= \ev_\alpha(h \cdot \F_q[x]^{< T}),
\end{alignat*}
where $\F_q \subseteq \F_q[x]$ is identified with the constant polynomials. If the evaluation points $\alpha$ are chosen such that $h(\alpha_i) \neq 0$, then these are all generalized Reed--Solomon codes. Thus, $d^\perp(\mathcal{C}_\ell^\noise) - 1 = X$ and $d^\perp(\mathcal{D}_\ell^\noise) - 1 = T$, which means that we achieve the desired security and privacy levels. Furthermore, it is clear that $\dim(\mathcal{C}_\ell) = \dim(\mathcal{D}_\ell) = 1$. The codes $\mathcal{E}_\ell$, $\mathcal{E}$ and $\mathcal{E}^\noise$ are
\begin{equation*}
    \mathcal{E}_\ell = \ev_\alpha(h_\ell \cdot \F_q), \quad \mathcal{E} = \ev_\alpha(\F_q[x]^{< L}), \quad \mathcal{E}^\noise = \ev_\alpha(h \cdot \F_q[x]^{< X + T - 1 + \Delta}).
\end{equation*}
By setting $N = L + X + T - 1 + \Delta$ we can guarantee that $\ev_\alpha$ is injective on all of the above subspaces, which means that $\mathcal{E} \cap \mathcal{E}^\noise = 0$ due to \cref{eq:truncated_polynomial_division_decomposition}. Therefore, according to \cref{thm:PIR_from_secret_sharing} there is an $X$-secure $T$-private information retrieval scheme with rate $\mathcal{R} = \tfrac{L}{N}$, where $N = L + X + T - 1 + \Delta$.

Recall that we want to choose $h_1, \dots, h_L$ to be a basis of $\F_q[x]^{< L}$ and $h$ of degree $L$ to be a multiple of $h_\ell$ for all $\ell \in [L]$, with the goal of minimizing $\Delta$. By choosing the basis polynomials $h_\ell = x^{\ell - 1}$ and $h = x^L$, we have that $\deg(h/h_\ell) = L - \ell + 1$, so $\Delta = L$, which scales with $L$. A better choice of basis elements can be made by choosing distinct elements $\gamma_1, \dots, \gamma_L \in \F_q$ and setting
\begin{equation*}
    h_\ell = \prod_{\substack{\ell' \in [L] \\ \ell' \neq \ell}} (x - \gamma_{\ell'}), \quad \text{and} \quad h = \prod_{\ell \in [L]} (x - \gamma_\ell).
\end{equation*}
We have that $h / h_\ell = x - \gamma_\ell$, so $\Delta = 1$. Therefore, $N = L + X + T$, which corresponds to the decomposition
\begin{equation}
    \F_q[x]^{< L} \oplus h \cdot \F_q[x]^{< X + T} = \F_q[x]^{< L + X + T} \label{eq:genus_zero_polynomial_decomposition}
\end{equation}
of finite dimensional vector spaces. We obtain the PIR rate $\mathcal{R} = 1- \tfrac{X + T}{N}$. This is the construction given in \cite{Jia_Sun_Jafar_XSTPIR} translated to polynomials. As we need to choose $N$ distinct evaluation points from $\F_q$ distinct from the $\gamma_1, \dots, \gamma_L$ we chose as the roots of $h$, we need the field size to be $q \geq N + L$. We recap this construction in the following theorem.

\begin{theorem}\label{thm:projective_line_construction}
Let $N = L + X + T$. If $q \geq L + T$, then there exists a PIR scheme over $\F_q$ with rate
\begin{equation*}
    \mathcal{R} = 1 - \frac{X + T}{N},
\end{equation*}
which is $X$-secure and $T$-private.
\end{theorem}

\subsection{Geometric Interpretation}

Let $\mathcal X = \mathbb{P}^1$ and let $P_\infty = [0:1]$ be the point at infinity.  We may write \eqref{eq:genus_zero_polynomial_decomposition} in terms of Riemann--Roch spaces as follows. Recall that $\mathcal{L}((k - 1)P_\infty) = \F_q[x]^{< k}$ and $h \cdot \mathcal{L}(D) = \mathcal{L}(D - (h))$. Therefore, \eqref{eq:genus_zero_polynomial_decomposition} corresponds to
\begin{equation}\label{eq:poly_decomposition_riemann_roch}
    \mathcal{L}((L - 1)P_\infty) \oplus \mathcal{L}((X + T - 1)P_\infty - (h)) = \mathcal{L}((L + X + T - 1)P_\infty),
\end{equation}
where $(h) = P_1 + \dots + P_L - LP_\infty$, and $P_\ell = [\gamma_\ell:1]$.  It is this direct sum decomposition of Riemann--Roch spaces that will generalize to higher genus curves, allowing for analogous PIR protocols from AG codes.  In the previous subsection we saw that choosing an appropriate basis of the space $\mathcal L((L-1)P_\infty)$ was crucial to the CSA construction.  Finding similar bases of analogous Riemann--Roch spaces for higher-genus curves will prove equally crucial.

\section{Construction Using Hyperelliptic Curves}\label{sec:hyperelliptic_curve_construction}

In this section we take the geometric interpretation for the polynomial construction of CSA codes of the previous section, and generalize it from the projective line to hyperelliptic curves.  Ultimately, this allows us to construct PIR protocols which operate on smaller fields than are possible with the genus zero construction of \cite{Jia_Sun_Jafar_XSTPIR}.  In the same way that Shamir secret sharing is an essential building block of the CSA codes of the previous section, underlying the construction of the current section is the Chen--Cramer secret sharing scheme of \cite{chen_ag_secret_share}.

Throughout this section we fix a hyperelliptic curve $\mathcal{X}$ defined by $y^2 + H(x)y = F(x)$ over $\F_q$ with genus $g \geq 1$, $\deg(F) = 2g + 1$ and $\deg(H) \leq g + 1$, and a single rational point $P_\infty$ at infinity.  Throughout this section, we assume that $L \equiv g \pmod 2$ and $L \geq g$ and set $J = \tfrac{L + g}{2}$.

\subsection{Information and Noise Spaces} \label{sec:info_noise}

The information space $\F_q[x]^{< L}$ from the previous section will be replaced with the subspace $\mathcal{L}((L + g - 1)P_\infty)$, which is also $L$-dimensional.  We let $h$ be a rational function on $\mathcal X$ with pole divisor $(h)_\infty = (L+g)P_\infty$.  If $U = \mathcal X\setminus \mathcal S$, where $\mathcal S$ is some finite subset of $\mathcal X$ containing $P_\infty$, then we have the containment
\begin{equation}\label{eq:higher_genus_direct_sum}
\mathcal L((L+g-1)P_\infty) \oplus h\cdot \mathcal O_{\mathcal X}(U) \subseteq \mathcal O_{\mathcal X}(U)
\end{equation}
which will provide us with the corresponding codes $\mathcal E$ and $\mathcal E^\noise$ as before, after restricting to appropriate finite-dimensional subspaces.

In analogy with the genus zero case, we have scalars $f_\ell, g_\ell\in \F_q$ whose product we want to compute.  We choose the secret shares to be
\begin{equation}\label{eq:hyperelliptic_curve_secret_shares}
    \hat{f}_\ell \in f_\ell + f_\ell^\noise \cdot \mathcal L(X + 2g - 1)P_\infty, \quad \hat{g}_\ell \in g_\ell h_\ell + g_\ell^\noise \cdot \mathcal L(T + 2g - 1)P_\infty
\end{equation}
for some rational functions $h_\ell$ which form a basis of $\mathcal L((L+g-1)P_\infty)$.  Again, we choose $g_\ell^\noise = h$ and $f_\ell^\noise = h/h_\ell$ and a short computation yields that
\[
\sum_{\ell \in [L]}\hat{f}_\ell\hat{g}_\ell \in \sum_{\ell\in[L]}f_\ell g_\ell h_\ell + h\cdot \mathcal L((X+T+4g-2)P_\infty + D_\infty)
\]
where $D_\infty = \max_\ell \{(h/h_\ell)_\infty\}$.  Let $\Delta = \deg(D_\infty)$, so that the dimension of our noise space is
\[
\ell((X+T+4g-2)P_\infty + D_\infty) = X + T + 3g - 1 + \Delta
\]
Our goal now is to choose $h$ and $h_\ell$ to attempt to minimize the above quantity.

To construct the basis functions and the noise space we choose distinct values $\gamma_1, \dots, \gamma_J \in \F_q$ such that $F(\gamma_j) \neq 0$ and set $h = \prod_{j \in [J]} (x - \gamma_j)$. Further, set
\begin{equation*}
    h_j^{(1)} = \prod_{\substack{j' \in [J] \\ j' \neq j}} (x - \gamma_{j'})~~\text{for $j \in [J]$}, \quad h_j^{(2)} = y \prod_{\substack{j' \in [J-g] \\ j' \neq j}} (x - \gamma_{j'})~~\text{for $j \in [J - g]$}.
\end{equation*}
It is simple to check that $h_1^{(1)}, \dots, h_J^{(1)}, h_1^{(2)}, \dots, h_{J-g}^{(2)}$ is a basis for $\mathcal{L}((L + g - 1)P_\infty)$. There are $L = 2J - g$ of these basis functions that we will denote by $h_1, \dots, h_L$. We have that
\begin{align*}
    h / h_j^{(1)} &= x - \gamma_j \in \mathcal{L}(2P_\infty) \\
    h / h_j^{(2)} &= y^{-1}(x - \gamma_j)(x - \gamma_{J - g + 1}) \cdots (x - \gamma_J) \in \mathcal{L}(P_\infty + (y)_0).
\end{align*}
It follows that $D_\infty = 2P_\infty + (y)_0$ and therefore $\Delta = \deg(2P_\infty + (y)_0) = 2g + 3$.  Hence the dimension of the noise space is $\ell((X+T+4g)P_\infty + (y)_0) = X + T + 5g + 2$.

Observe that nonzero functions in the noise space $h\cdot \mathcal L((X+T+4g)P_\infty + (y)_0)$ have at least $2J = L + g$ zeros, but nonzero functions in the information space $\mathcal L((L+g-1)P_\infty)$ can have at most $L + g - 1$ zeros. Therefore, the intersection of these two Riemann--Roch spaces is trivial.  Considering the smallest Riemann--Roch space that contains both the information and noise spaces, we have the inclusion
\begin{equation*}
    \mathcal L((L+g-1)P_\infty) \oplus h\cdot \mathcal L((X+T+4g)P_\infty + (y)_0)\subseteq \mathcal{L}((L + X + T + 5g)P_\infty + (y)_0)
\end{equation*}
which is the higher genus analogue of the direct sum decomposition of \eqref{eq:poly_decomposition_riemann_roch}.  Setting $\mathcal S = \{P_\infty\}\cup \supp((y)_0)$, we also see that the above direct sum is a subspace of the direct sum appearing in \eqref{eq:higher_genus_direct_sum}, as was the case with the analogous objects in genus zero.

\subsection{The Evaluation Map and PIR Rate}

Let us now finalize our construction by choosing a set of evaluation points and computing the PIR rate.  With the objects and notation of the previous subsection, we denote $D = (L + X + T + 5g)P_\infty + (y)_0$. By finding $\deg(D) + 1 = L + X + T + 7g + 2$ rational points $P_n$ distinct from $P_\infty$ and $(y)_0$ such that $h(P_n) \neq 0$, we can define the evaluation map and the associated AG code. The dimension of the resulting code will be $N = \deg(D) + 1 - g = L + X + T + 6g + 2$. By restricting the code to an information set $\mathcal P$ of size $N$ the code we get an injective evaluation map $\ev_\mathcal{P}$. The corresponding secret sharing codes, which in turn define our PIR scheme, are then
\begin{alignat*}{2}
    \mathcal{C}_\ell &= \ev_\mathcal{P}(\F_q), &\qquad \mathcal{C}_\ell^\noise &= \ev_\mathcal{P}(h/h_\ell \cdot \mathcal{L}((X + 2g - 1)P_\infty)) \\
    \mathcal{D}_\ell &= \ev_\mathcal{P}(h_\ell \cdot \F_q), & \mathcal{D}_\ell^\noise &= \ev_\mathcal{P}(h \cdot \mathcal{L}((T + 2g - 1)P_\infty),
\end{alignat*}
where $\F_q \subseteq \F_q(\mathcal{X})$ is identified with the constants. According to \cref{eq:AG_dual_distance}, $d^\perp(\mathcal{C}_\ell^\noise) - 1 \geq X$ and $d^\perp(\mathcal{D}_\ell^\noise) - 1 \geq T$, which means that we achieve the desired security and privacy levels. It is again clear that $\dim(\mathcal{C}_\ell) = \dim(\mathcal{D}_\ell) = 1$. The resulting codes $\mathcal{E}_\ell, \mathcal{E}$ and $\mathcal{E}^\noise$ are
\begin{align*}
    \mathcal{E}_\ell &= \ev_\mathcal{P}(h_\ell \cdot \F_q), \quad \mathcal{E} = \ev_\mathcal{P}(\mathcal{L}((L + g - 1)P_\infty)) \\
    \mathcal{E}^\noise &= \ev_\mathcal{P}(h \cdot \mathcal{L}((X + T + 4g)P_\infty + (y)_0)).
\end{align*}
Finally, we arrive at the following theorem.

\begin{theorem}\label{thm:AG_construction}
Let $\mathcal{X}$ be a hyperelliptic curve of genus $g$ defined by the affine equation $y^2 + H(x)y = F(x)$ over $\F_q$. Let $L \geq g$, $L \equiv g \pmod 2$ and $J = \tfrac{L + g}{2}$. Let $\gamma_1, \dots, \gamma_J \in \F_q$ be such that $F(\gamma_j) \neq 0$ and set $h = \prod_{j \in [J]} (x - \gamma_j)$. Let $N = L + X + T + 6g + 2$ for some security and privacy parameters $X$ and $T$. If there are $N + g$ rational points $P_1, \dots, P_{N + g}$ disjoint from $P_\infty$ and $(y)_0$ and such that $h(P_n) \neq 0$, then there exists a PIR scheme over $\F_q$ with rate
\begin{equation*}
    \mathcal{R} = 1 - \frac{X + T + 6g + 2}{N}.
\end{equation*}
which is $X$-secure and $T$-private.
\end{theorem}

\subsection{Maximizing the Rate}

Instead of focusing on the PIR rate given some fixed number of servers $N$, we want to fix a field size $q$ and find the maximal rate over that field size given the security and privacy parameters $X$ and $T$. For the construction over the projective line we require that $q \geq 2L + X + T$, so the largest rate is achieved by choosing $L = \lfloor \tfrac{q - (X + T)}{2} \rfloor$ and $N = L + X + T$.

For a given hyperelliptic curve $\mathcal{X}$ of genus $g$ the maximum PIR rate is achieved by maximizing $N$, i.e., by maximizing the parameter $L$. Recall that we have to be able to find $N + g$ rational points that are disjoint from $\{P_\infty\} \cup \supp((y)_0) \cup \supp((h)_0)$. The function $h$ depends on $L$, so we want to maximize $L$, while keeping the number of rational points in $\supp((h)_0)$ sufficiently small. The zeros of $h$ are those points whose $x$-coordinate is $\gamma_j \in \F_q$ for $j = 1, \dots, J$.

Let $\overline{\Gamma} = \{x(P) \mid P \in \mathcal{X}(\F_q) \}$ and $\Gamma = \F_q \setminus \overline{\Gamma}$. If $\gamma \in \Gamma$, then $x - \gamma$ has no rational zeros and if $\gamma \in \overline{\Gamma}$, then $x - \gamma$ has at most two rational zeros. This means that choosing $\gamma_j \in \Gamma$ will not reduce the number of rational points at our disposal, but choosing $\gamma_j \in \overline{\Gamma}$ will reduce the number by at most two. Choosing an additional $\gamma_j$, i.e., increasing $J$ by one, will increase $L$ by two. We will choose the $\gamma_j$'s by first choosing them from $\Gamma$ and then from $\overline{\Gamma}$ until we cannot choose any more points without violating the condition of having at least $N + g$ rational points outside of $\{P_\infty\} \cup \supp((y)_0) \cup \supp((h)_0)$. While it is hard to derive a precise expression for the resulting $L$, the following example will show that choosing a maximal curve will not always yield the best PIR rate.

\begin{example}\label{ex:maximal_curve_not_optimal}
Set $X = T = 1$ and consider curves of genus $g = 1$. Consider the maximal elliptic curve over $\F_{11}$ defined by $y^2 = x^3 + x + 3$ with $18$ rational points. The function $y$ has one rational zero, which means that the number of usable rational points is $16$. We see that the number of free $x$ coordinates is $\lvert \Gamma \rvert = 2$. We choose $\gamma_1, \gamma_2 \in \Gamma$ and set $J = 2, L = 3$. This implies that $N = 13$, so we need to have $N + g = 14$ rational points available outside of $P_\infty$ and the rational zero of $y$. As $18 - 2 = 16 \geq 14$, we see that having $J = 2$ is possible. On the other hand, setting $J = 3$ and choosing $\gamma_3 \in \overline{\Gamma}$ is not possible, since this reduces the number of usable rational points to $14$, while increasing the requirement to $N + 1 = 16$. Therefore, the maximal achievable rate for this curve is $\tfrac{3}{13} \approx 0.231$.

Consider the elliptic curve over $\F_{11}$ defined by $y^2 = x^3 + 2x + 4$ with $17$ rational points. The function $y$ has no rational zeros, which means that there are again $16$ usable rational points. The number of free $x$ coordinates is $\lvert \Gamma \rvert = 3$, so let us set $J = 3, L = 5$ and choose $\gamma_1, \gamma_2, \gamma_3 \in \Gamma$. This means that $N = 15$ and we achieve a rate of $\tfrac{5}{15} \approx 0.333$.
\end{example}

\section{Discussion}\label{sec:discussion}

In this section we discuss the different aspects our construction and compare it to the CSA construction in \cite{Jia_Sun_Jafar_XSTPIR}.

One of the difficulties in our construction is finding hyperelliptic curves with a large number of rational points.  Exhaustive search can be done for small values of $g$ and $q$ by going through all polynomials $F(x), H(x)$ in the equation of the hyperelliptic curve.  For larger parameters, such an exhaustive search is not feasible, but random search over $F(x), H(x)$ does produce curves with sufficiently many points to demonstrate the desired improvement in the rate as the genus increases.  Lastly, for some values of $g$ and $q$, hyperelliptic curves in the desired form with many points can be found at \cite{many_points}.

\subsection{Comparison}
\begin{figure}
    \centering
    \includegraphics[width=\textwidth]{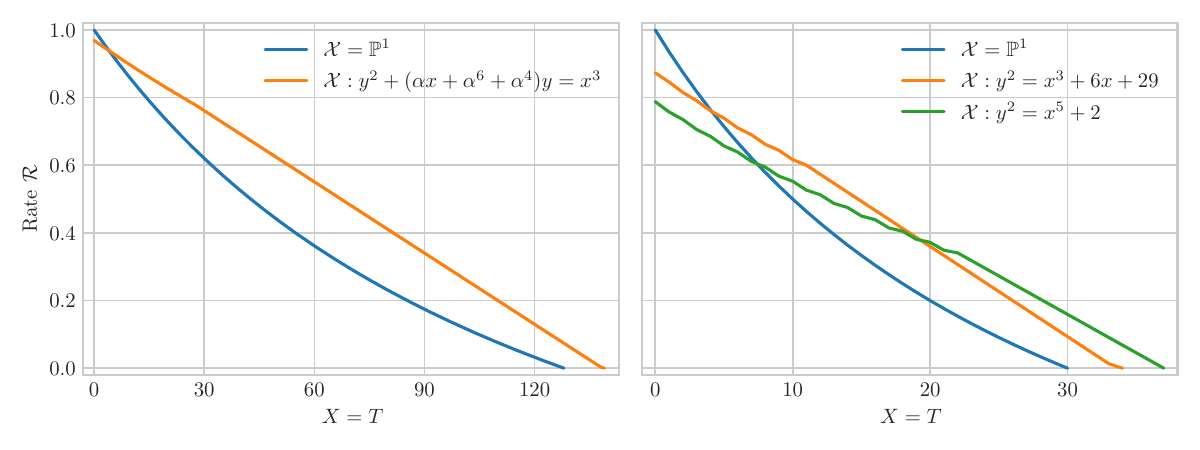}
    \label{fig:comparison}
    \caption{Comparison between the maximal achievable rate of the CSA construction of \cite{Jia_Sun_Jafar_XSTPIR} over the projective line (\cref{thm:projective_line_construction}) and the construction over hyperelliptic curves (\cref{thm:AG_construction}) over a fixed field. The constructions are over $\F_{2^8}$ with primitive element $\alpha$ satisfying $\alpha^8 + \alpha^4 + \alpha^3 + \alpha^2 + 1 = 0$ (left) and over $\F_{61}$ (right).}
\end{figure}

As mentioned before, we are interested in the maximal PIR rate achievable over a fixed field size $q$ and fixed security and privacy parameters $X$ and $T$. In \cref{fig:comparison} we have plotted the maximal rate as a function of $X = T$ over different curves. We observe that the constructions over hyperelliptic curves achieve a higher maximal PIR rate when $X = T$ is sufficiently large. The following example shows how the flexibility in choosing a larger parameter $L$ allows for the larger rate.

\begin{example}
Consider the field $\F_{2^8}$, which can be represented in one byte and has efficient hardware implementations on many architectures. Set the security and privacy parameters to $X = T = 50$. Over this field, the largest rate achievable for the construction over the projective line is achieved for $L = 78$ and $N = 178$, since this uses all $256 = 2 \cdot 78 + 50 + 50$ points. The corresponding PIR rate is $\mathcal{R} = \tfrac{78}{178} \approx 0.44$. On the other hand, consider the elliptic curve $\mathcal{X} : y^2 + (\alpha x + \alpha^6 + \alpha^4)y = x^3$ with 288 rational points, where $\alpha \in \F_{2^8}$ satisfies $\alpha^8 + \alpha^4 + \alpha^3 + \alpha^2 + 1 = 0$. This construction allows for $L = 177$ and $N = 285$ with a corresponding PIR rate of $\mathcal{R} = \tfrac{177}{285} \approx 0.62$. This shows that over the same field $\F_{2^8}$ the construction given in this paper can achieve a much higher PIR rate.
\end{example}

\subsection{Security and Privacy Beyond $X$ and $T$}

Let us now briefly highlight an additional feature of our PIR scheme, namely that it is private against some $U$-subsets of servers even when $U > T$, and similarly secure against some $V$-subsets of servers even when $V > X$.  As we will see, the CSA codes of \cite{Jia_Sun_Jafar_XSTPIR} cannot have this property, because the underlying LSSS is based on MDS codes.

Returning to the language of LSSS's of \cref{sec:secret_sharing}, one has for a general code $\mathcal C^\noise$ that $d^\perp(\mathcal C^\noise) - 1 \leq \dim(\mathcal C^\noise)$, with equality exactly when $\mathcal C^\noise$ is MDS. If $(\mathcal C, \mathcal C^\noise)$ is a $T$-secure LSSS, then we can consider its security against sets of $U$ compromised parties for $U\geq T$. Clearly, the interval of interest is
\begin{equation}\label{eq:security_interval}
U \in [d^\perp(\mathcal C^\noise) - 1, \dim(\mathcal C^\noise)]
\end{equation}
on which \cref{prop:secret_sharing_security} admits the following generalization.

\begin{proposition}\label{prop:secret_sharing_flexible_security}
    Let $(\mathcal C,\mathcal C^\noise)$ be an LSSS and let $\mathcal T\subseteq[N]$ be a subset of size $U\in [d^\perp(\mathcal C^\noise) - 1, \dim(\mathcal C^\noise)]$. If $\dim(\mathcal C^\noise(\mathcal T)) = U$, then the LSSS is secure against $\mathcal T$.
\end{proposition}

For $U$ in the interval of \eqref{eq:security_interval} we define $\sigma(U)$ to be the fraction of $U$-subsets of $[N]$ against which an LSSS is \emph{not} secure.  If we consider a Chen--Cramer LSSS with $\mathcal C^\noise = \mathcal C(\mathcal P, (T+2g-1)P_\infty-(h))$ as in \cref{ex:chen_cramer_secret_sharing}, then we are secure against some $U$-sets whenever $U \in [T, T+g]$.  The authors of \cite{ag_secret_share_asymptotic} study the function $\sigma(U)$ on this interval and show that if $g/q^{1/2} \to 0$ then $\sigma(U)\to 0$ for all $U\in [T,T+g]$.  For the $[24, 6, 17]$ AG code over $\F_{13}$ from \cref{ex:chen_cramer_secret_sharing}, one can compute by enumerating low weight codewords of the dual code that $\sigma(5) = 92/\binom{24}{5} \approx 0.0022$ and $\sigma(6) = 8684/\binom{24}{6} \approx 0.0645$.

As the security and privacy properties of our PIR scheme are inherited directly from the underlying LSSS's, it is clear that the PIR scheme of \cref{sec:hyperelliptic_curve_construction} is secure against some subsets of size $V\in [X,X+g]$, and private against some subsets of size $U\in [T,T+g]$.  However, we are most interested in the case of a fixed ground field, the asymptotic results of \cite{ag_secret_share_asymptotic} are not directly applicable to the current work.  While isolated examples such as that of the previous paragraph hint that the related quantities $\sigma(U)$ are indeed quite small, we reserve deeper study of the security and privacy properties of PIR schemes beyond the prescribed parameters $X$ and $T$ for future work.

\subsection{Conclusions and Future Work}

In this  work, we have reformulated the original cross-subspace alignment scheme for secure PIR in the language of Reed--Solomon codes on the projective line, and generalized the construction to utilize algebraic geometry codes on hyperelliptic curves of arbitrary genus. A higher genus yields more rational points with respect to the field size, hence allowing for a more flexible choice of parameters. For instance, for a fixed field size, we can increase the number of servers beyond the field size, which is not possible for genus zero. This enables higher rates at the cost of slightly higher subpacketization level and few more servers. By the new construction, we can also avoid sharp threshold effects that may lead to suboptimal implementations. For instance, arithmetic over certain finite fields is highly optimized on hardware. Nevertheless, for application settings where the field size or computational complexity is not a concern, the original CSA codes employed with large enough fields provide higher rates.

As for future work, one could consider establishing upper bounds on the maximal rate achievable by our construction, that is, for a fixed field size $q$ and genus $g$, find explicit upper bounds on the PIR rate as we range over all hyperelliptic curves of the given genus and all possible choices of interpolation and evaluation points.  More generally, one could ask for upper bounds on the achievable secure PIR rate for a fixed field size, though such questions are notoriously difficult.  Another interesting direction for future research is to study similar constructions for other families of algebraic curves, including Hermitian or norm-trace curves. This means that one should find a suitable basis of $\mathcal{L}((L + g - 1)P_\infty)$, or some other Riemann--Roch space, that achieves a constant $\Delta$-parameter (as described in \cref{sec:info_noise}) with respect to $L$. Constructions over other families of curves would also clarify the relevant properties that are required from the curves used in homomorphic secret sharing.

\bibliography{bib.bib}
\bibliographystyle{ieeetr}

\end{document}